\documentclass[%
 reprint,
showpacs,preprintnumbers,
 amsmath,amssymb,
 aps,
prl,
]{revtex4-1}

\usepackage{graphicx}
\usepackage{dcolumn}
\usepackage{bm}
\usepackage{epstopdf}

\begin{document}


\title{Clear Identification of Previously Unresolved Overlapping Resonances in Dissociative Electron Attachment: A Case Study on Chlorine Molecule}

\author{Pamir Nag}
\author{Dhananjay Nandi}%
 \email{dhananjay@iiserkol.ac.in}
\affiliation{Indian Institute of Science Education and Research (IISER)-Kolkata, Mohanpur -741246, India
}%




\date{\today}
\begin{abstract}
An observed broad resonance peak in dissociative electron attachment (DEA) cross sections to a molecule might be due to either closely lying or overlapping resonances involved in the process. We developed a state-of-the-art experimental approach that enables us to identify unambiguously both the closely lying and the overlapping resonances based on kinetic energy and/or angular distribution measurements of fragment negative ion(s) generated from DEA process. Experimental observations strongly supported by theoretical potential energy curve calculations unambiguously identify previously unresolved overlapping resonances in DEA to chlorine molecule.

\end{abstract}

\maketitle

Dissociative electron attachment (DEA) is in general a two step process that happens in almost all branches of science.
In the first step, the electron is captured resonantly by the molecule forming temporary negative ion (TNI) state that is in general unstable and repulsive in nature. The TNI subsequently dissociates into final products of fragment negative ion(s) and neutral(s). The DEA process is a nice tool to study resonances and dynamics occurs in the electron-molecule collisions. The resonances are generally studied by measuring cross sectional curve of fragment negative ion(s), whereas for dynamics the kinetic energy and angular distributions are typically probed. Closely lying resonances in DEA process may result a broad peak or structures in the cross sectional curve depending on the resolution of the primary electron beam. In order to separate out the resonances, it is absolutely necessary to use primary electron beam having energy resolution better than the peak energy difference of the resonances involved. Whereas, the overlapping resonances always show a broad resonant peak no matter how good is the energy resolution of the primary beam.  

In this Letter we show an unique way to identify overlapping resonances even if the energy resolution of the primary electron beam is far worse than the peak energy difference of the resonances involved. The identification is done based on the angular distribution measurements of fragment negative ion. Here we demonstrate our methodology taking a benchmark experiment: probing the previously unresolved resonance at 5.7 eV in DEA cross section curve to chlorine (Cl$_{2}$) molecule. Experimental observation is strongly supported by {\it ab initio} potential energy curve calculation using density functional theory (DFT).   
 
Three resonant peaks were observed in DEA to Cl$_{2}$ in all previous experimental studies \cite{cl:kurepa,cl:azria,cl:mark}: near 0 eV, 2.5 eV and a broad peak around 5.7 eV.  An additional peak at $9.7\pm0.2$ eV was also observed only by Kurepa and Beli{\'c} \cite{cl:kurepa} and assigned as a $^{2}\Sigma_{g}^{+}$ TNI state. The first two resonances were well studied both theoretically \cite{cl:thierry,cl:fabrikant,cl:tasker} and experimentally \cite{cl:prl,cl:mason} using variety of techniques and conclusively identified a $^{2}\Sigma_{u}^{+}$ and a $^{2}\Pi_{g}$ TNI state for each resonance, respectively. 
In principle, different resonances can be experimentally identified either by measuring the DEA cross section with high resolution electron beam or by measuring the kinetic energy of the fragment negative ions as a function of incident electron energy. However, in the recent absolute cross section measurement \cite{cl:mark} using high resolution ($\sim$ 80 meV) electron beam also observed a broad resonant peak around 5.5 eV. From the ion yield curve in order of increasing energy they assigned a $^2\Pi_u$ TNI state for this resonance as the $^2\Pi_u$ state lies below the $^2\Sigma_g^+$ state theoretically.
No structure around the resonance implies if at all two resonances are present then the corresponding TNI states are very closely lying in the Franck-Condon transition region and the resonances are heavily overlapped.

Two overlapping resonances can also be identified from the angular distribution measurements of fragment negative ion. In fact, Azria \emph{et al.} \cite{cl:azria} measured the angular distribution of Cl$^-$ fragment using conventional technique over a limited angular range and concluded that the observed structure in the Cl$^{-}$ ion yield curve in the energy range 4-8 eV could be due to either (i) a change with energy of the mixing of two partial waves in single resonance ($^{2}\Pi_{u}$) TNI state or (ii) to a small contribution of the $^{2}\Sigma_{g}^{+}$ state below the $^{2}\Pi_{u}$ resonant state. But the $^2\Sigma_g^+$ state is always above the $^2\Pi_u$ state in all theoretically computed potential energy curves. Moreover, the computed $^2\Pi_u$ and $^2\Sigma_g^+$ states \cite{cl:gilbert} are out of Franck-Condon overlap region in the measured energy range. Thus, the open questions are: (i) convergence between the theory and experiment yet to be achieved and (ii) the reason for having such a broad resonance peak around 5.7 eV yet to be resolved. 

\begin{figure}[b]
\centering
\includegraphics[scale=0.4]{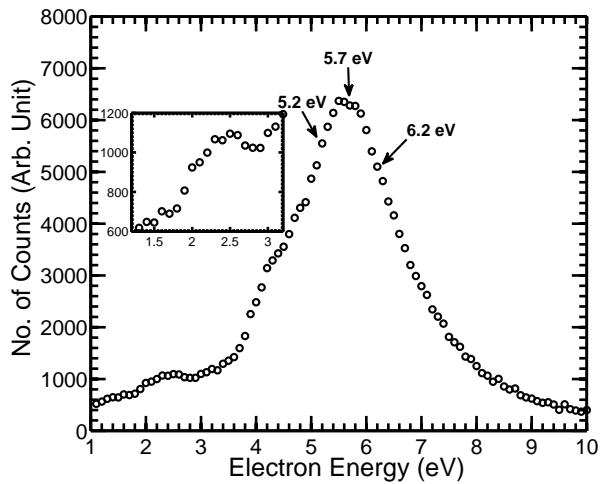}
\caption{Ion yield curve of Cl$^-$ arising from DEA to Cl$_2$ over a limited energy range. The arrows indicates the energies at which the velocity slice images are taken.} \label{fig:ion_yield}
\end{figure}

We propose and provide a clear evidence of the presence of two overlapping resonances around the 5.7 eV peak of Cl$^-$ cross sectional curve arising from DEA to Cl$_2$. By measuring the angular distribution over 2$\pi$ angular range using currently developed highly-differential state-of-the-art velocity slice imaging (VSI) technique and from computed potential energy curve using density functional theory (DFT) the presence of two overlapping resonances have been clearly identified. Here, our particular emphasis is on the symmetry determination of Cl$_{2}^{-}$ resonant state(s) responsible for the broad peak in the yield curve of Cl$^{-}$ formation around 5.7 eV. In our earlier studies \cite{ar:nag15,NO:DN}, we could clearly identify two closely lying resonances from the analysis of both kinetic energy and angular distribution data. But, presently studied resonance seems to be heavily overlapped that we are unable to separate out from the kinetic energy distribution measurements. However, those overlapping resonances are clearly identified based on the angular distribution analysis. Involvement of a $^2\Pi_u$ and a  $^2\Sigma_g^+$ TNI states around this peak have been observed. The potential energy curves of neutral chlorine molecule and molecular negative ions using {\it ab initio} method have also been computed. All four TNI states that exactly matches with the experimentally observed resonant energies \cite{cl:mark} for the DEA to chlorine molecules could be reproduced.   

In the current experiment velocity map imaging technique \cite{ar:parker} in combination of slicing idea \cite{ar:ashfold} have been used to measure the kinetic energy distribution and angular distribution over the entire $2\pi$ angle simultaneously. The experimental setup is almost identical as described previously \cite{ar:nandi05} with minor modifications. In brief, the entire experiment is performed inside an oil-free ultra high vacuum chamber with a base pressure below $\sim$~10$^{-9}$ mbar to ensure single collision condition. An effusive molecular beam is crossed by a pulsed electron beam of typical energy resolution of 0.6 eV. The electron beam is generated by thermionic emission from a tungsten filament and magnetically collimated using about 35 Gauss magnetic field produced by a pair of coils in Helmholtz configuration mounted outside the chamber. The electron beam pulse is of 200 ns width and repeated with 10 kHz frequency. The velocity map imaging (VMI) spectrometer is a three field time-of-flight (ToF) spectrometer with a 100 mm long field free flight tube. The VMI spectrometer focuses the ions created with a given velocity in the interaction region to a single point on a 2-dimensional position sensitive detector (PSD) irrespective of the point of creation. Three micro channel plates (MCP) of 80 mm diameter in Z-stack configuration with hexanode \cite{hex1, hex2} from Roentdek is used as PSD which records the (x, y) position along with the time of flight (t) of each particle detected. A 4 $\mu$s long negative extraction pulse is applied after 100 ns of the electron gun pulse at the first electrode of the spectrometer to pulsed extract the ions. Due to this delay and also while traveling through the flight tube the ion clouds expand that is very helpful for better slicing. During the analysis a 50 ns thin slice is taken to select the central part of the Newton sphere that actually landed on the detector with biggest diameter. This central sliced image contains both the kinetic energy and angular distribution information.

\begin{figure*}
\centering
\includegraphics[scale=0.42]{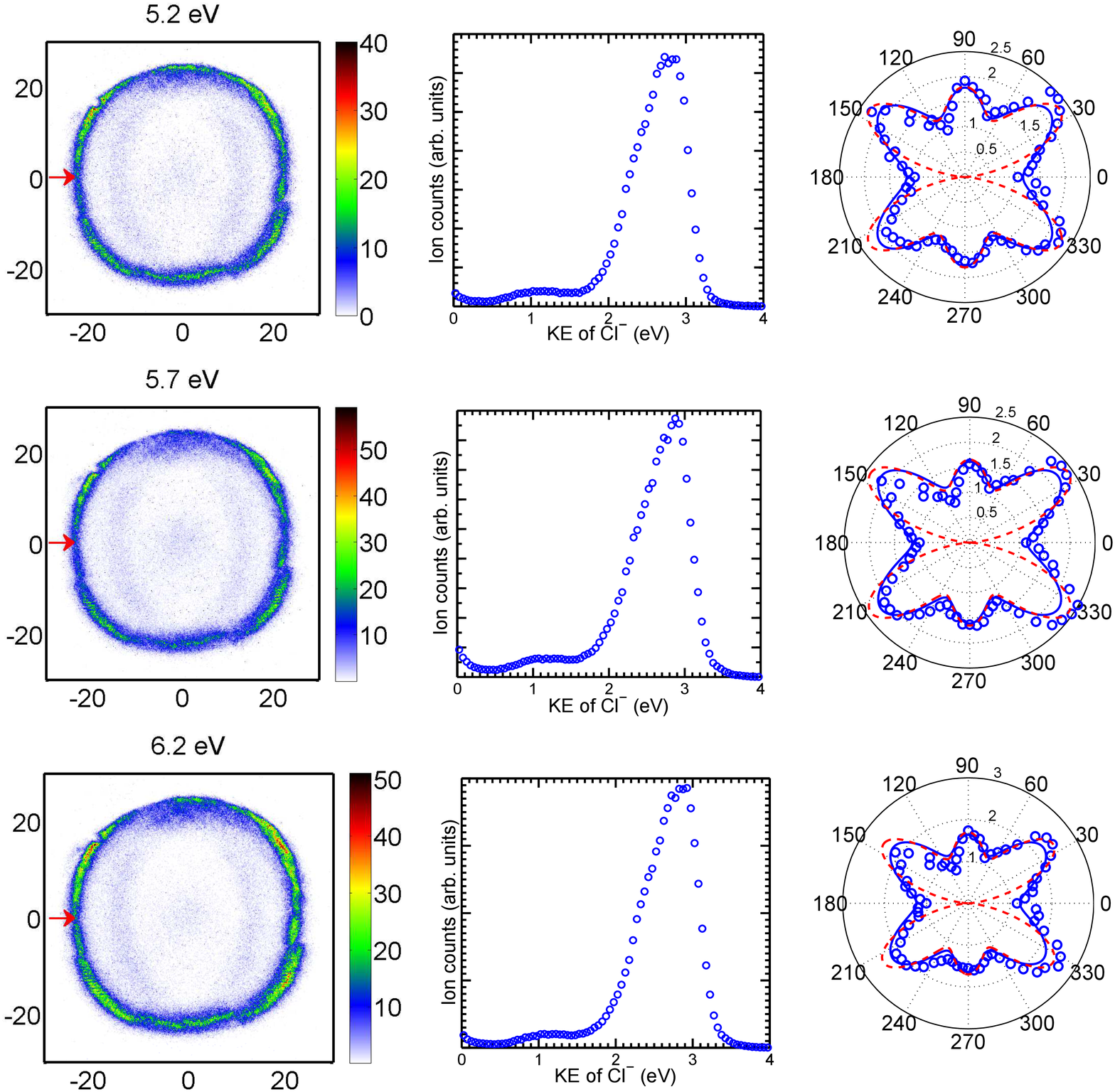}
\caption{(Color online) Velocity slice images (left), kinetic energy distribution (middle) and angular distributions (right) of Cl$^{-}$ ions arising from DEA to Cl$_{2}$ taken at resonance energy (middle), 0.5 eV below (top) and above (bottom) of the resonant energy, respectively. The electron beam direction is from left to right and through the center of each images as indicated by arrows.} \label{fig:ang}
\end{figure*}

The experiment is performed with 99.9\% pure commercially available chlorine gas. The electron beam energy calibration is performed using 6.5 eV DEA peak of O$^{-}$ from O$_{2}$ \cite{ref:rapp}. The ion yield curve over a limited energy range is shown in Fig.~\ref{fig:ion_yield}. This is in consistent with the recent report by Feketeova \emph{et al.} \cite{cl:mark}. A small peak at 2.5 eV and a broad resonance without any structure around 5.7 eV are observed, good agreement with all previous studies. Our main emphasis is on the broad resonance where we claim that two overlapping resonances are present. As already discussed, it is not yet possible to separate out the claimed two resonances either from cross section measurement or kinetic energy distribution. But, a clear signature of two resonances can be obtained from the angular distribution measurement of the Cl$^-$ ions. The central sliced images taken at the indicated incident electron energies around the studied resonance are shown in the first column of Fig.~\ref{fig:ang}. The arrows indicate the direction of the incident electron beam. Unlike our previous studies \cite{ar:nag15,NO:DN} we do not observe any noticeable change in the images as the incident electron energy changes. Close inspection to the sliced images show presence of some faint ring pattern with smaller diameter, we believe these are due to some chloride impurities created due to reaction with the highly reactive chlorine gas in the collision region and also in the gas line. The Cl$^-$ ions created from electron collision with Cl$_2$ also shows two closely lying ring patterns, these could be due to presence of two isotopes, e.g., $^{35}$Cl and $^{37}$Cl. To check that these are not an experimental artifact we have performed experiments with O$^-$/CO and O$^-$/O$_2$ without changing the experimental conditions before and after taking the images for Cl$^-$/Cl$_2$ and found the momentum distribution as expected. These sliced images contain both the kinetic energy and angular distribution information.

To determine the kinetic energy distribution of the Cl$^-$ ions from the sliced images, the spectrometer first calibrated with the kinetic energy release of O$^-$/O$_2$ \cite{o2:dn_cross} and considering the square of the radius of the image is proportional to the kinetic energy of the ions \cite{book:VMI}. The observed kinetic energy distributions from each images are shown in the corresponding middle column of Fig.~\ref{fig:ang}. As can be seen from each curve, three bands are observed in the distribution. The dominant peak in the kinetic energy distribution (around 2.5 eV, fairly matches with the value calculated from thermochemical parameters) is shifted as the electron energy increases as it is expected. The two lower energy bands are remained unchanged as the electron energy changes. These observation could support our previous claim that the dominating ring patter is coming from pure chlorine whereas the faint distributions with lower energies are from the impurities. From the kinetic energy distribution also it is not possible to separate out the proposed two overlapped resonances.

The angular distribution of the Cl$^{-}$ ions associated with the corresponding images are shown at the third column of Fig.~\ref{fig:ang}. All the Cl$^{-}$ ions created with kinetic energy in the dominating band (1.8 - 3.5 eV) are considered in the angular distribution measurement. The angular distributions peaking at around 40$^\circ$, 90$^\circ$ and 140$^\circ$ directions. A non-zero intensity is observed in the forward (0$^{\circ}$) and backward (180$^{\circ}$) directions. Being a homonuclear diatomic molecule no forward-backward asymmetry is observed as expected. From the angular distribution of the Cl$^{-}$ ions, the symmetry of the involved TNI states of Cl$_{2}^{-}$ has been determined using the expression derived by O' Malley and Taylor \cite{ang:omalley} and modified by Tronc \emph{et al.} \cite{ang:tronc}. During the analysis of the angular distribution the Dunn's selection rule \cite{ang:dunn} for homonuclear diatomic molecule are also taken care.

The ground state of Cl$_{2}$ molecule is $^{1}\Sigma_{g}^{+}$ ($\ldots, 5\sigma_{g}^{2}, 2\pi_{ux}^{2}, 2\pi_{uy}^{2}, 2\pi_{gx}^{2}, 2\pi_{gy}^{2}$). Like earlier studies, if we consider only a single TNI state with $\Pi_{u}$ symmetry is involved in the DEA process, the expected angular distribution should look like as presented with dashed (red) curve in the third column of the Fig~\ref{fig:ang}. Now, according to Dunn's selection rule \cite{ang:dunn}, for a $\Sigma_{g} \rightarrow \Pi_{u}$ transition, a zero intensity is expected in the 0$^{\circ}$ and 180$^{\circ}$ directions. A non-zero intensity in the forward and backward directions clearly indicates that only  a $\Pi_{u}$ state can not explain the experimentally observed angular distribution. Next accessible TNI state is a $^{2}\Sigma_{g} ^{+}$ within the energy range of interest and is slightly above the $^{2}\Pi_{u}$ state. In the Franck-Condon region, these two states are very close and both of them might be involved in the DEA process. This could be the reason for the observed broad peak in the cross section curve. The best fit to the data points is observed using the above mentioned two overlapping TNI states and displayed with solid (blue) curve in middle column of the Fig~\ref{fig:ang}. Present observations are in the line with the earlier experimental findings of Azria \emph{et al.} \cite{cl:azria} using angular distribution measurement in a limited angular range. To further support our claim, we performed potential energy curve calculation as described below.

\begin{figure}
\centering
\includegraphics[scale=0.44]{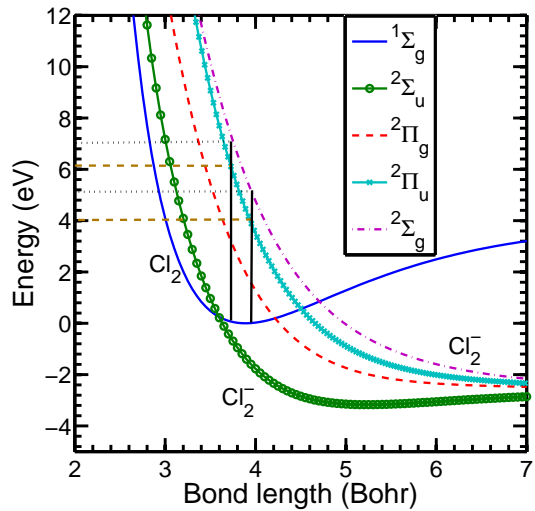}
\caption{(Color online) Potential energy curves of neutral ground state chlorine and few of its anionic states relevant for the DEA process and applying B3LYP functional with 6-311++G$^{**}$ basis set in the DFT calculation.} \label{fig:theory}
\end{figure}

We have computed the potential energy curve of neutral Cl$_{2}$ molecule in ground and different negative molecular ion states by density functional theory (DFT) calculation using MOLPRO package \cite{molpro1, molpro2}. The B3LYP functional with 6-311++G$^{**}$ basis set is used in the calculation. Being homonuclear diatomic molecule both Cl$_{2}$ and Cl$_{2}^{-}$ belong to $D_{\infty h}$ point group symmetry but MOLPRO uses only Abelian point group symmetry \cite{molpro:manual1} so $D_{2h}$ point group symmetry for the present case is used. The orbitals of chlorine molecule and molecular negative ions with symmetries belong to $D_{\infty h}$ point group are represented in terms of the symmetries of $D_{2h}$ point group as described in \cite{molpro:manual2}. The ground state of neutral Cl$_{2}$ molecule,  $^{1}\Sigma_{g}^{+}$ ($\ldots$, $5\sigma_{g}^{2}$,$2\pi_{ux}^{2}$,$2\pi_{uy}^{2}$,$2\pi_{gx}^{2}$,$2\pi_{gy}^{2} $), ground state of Cl$_{2}^{-}$ molecular ion $^{2}\Sigma_{u}$ with configuration $\ldots$, $2\pi_{gx}^{2}$,$2\pi_{gy}^{2}$,$5\sigma_{u}^{1}$ and three more excited states of Cl$_{2}^{-}$ having a single hole in close shell configuration $^{2}\Pi_{g}$ ($\ldots,2\pi_{gx}^{2}$,$2\pi_{gy}^{1}$, $5\sigma_{u}^{2}$), $^{2}\Pi_{u}$ ($\ldots, 5\sigma_{g}^{2}$,$2\pi_{ux}^{1}$,$2\pi_{uy}^{2}$,$2\pi_{gx}^{2}$,$2\pi_{gy}^{2}$, $5\sigma_{u}^{2}$) and $^{2}\Sigma_{g}$ ($\ldots, 5\sigma_{g}^{1}$,$2\pi_{ux}^{2}$,$2\pi_{uy}^{2}$,$2\pi_{gx}^{2}$,$2\pi_{gy}^{2}$,$5\sigma_{u}^{2}$) are computed over a wide energy range and shown in Fig.~\ref{fig:theory}. From the potential energy curve it evident that in Franck-Condon overlap region around 5.7 eV energy difference both $^{2}\Pi_{u}$ and $^{2}\Sigma_{g}$ states are presents and may be involved in the DEA process. The Franck-Condon overlap region with $^{2}\Pi_{g}$ starts around 4 eV and continues till nearly 6 eV whereas the overlap with $^{2}\Sigma_{g}$ starts around 5 eV and continues till nearly 7 eV. The overlapped region is shown using vertical lines in the Fig.~\ref{fig:theory}. This computed potential energy curves can successfully described the observed broad resonance staring around 4 eV and continues upto nearly 7.5 eV in the measured cross section \cite{cl:mark} and also strongly support our claim of the involvement of two resonance states.

In summary, we have shown even if the difference in the peak energy of two overlapping resonances is much smaller than the energy resolution of the primary electron beam, it is still possible to clearly identify from the angular distribution measurements of fragment negative ion arising from DEA to molecule of interest. Here, we demonstrate a case study by probing DEA to Cl$_{2}$, an excellent candidate, using VSI technique to justify our claim. The analysis of measured angular distribution confirms the presence of two overlapping resonances that explains the observed broad peak in Cl$^{-}$ cross section curve. The experimental observations are strongly supported by the potential energy curve calculation using DFT. The discussed two states are in the Franck-Condon overlap region and are excellent  match with the experimentally observed energy. These findings actually resolve the previous disagreement between theory and experimental results for the near 5.7 eV resonance peak of chlorine molecule.

D. N. gratefully acknowledges the partial financial support from ``Indian National Science Academy" for the development of VSI spectrometer under INSA Young Scientist project ``SP/YSP/80/2013/734".

\bibliography{clbib}

\end{document}